# Structural, elastic, electronic, bonding, thermo-mechanical and optical properties of predicted NbAlB MAB phase in comparison to MoAlB: DFT based ab-initio insights

Mst. Bina Aktar, F. Parvin, A. K. M. Azharul Islam, S. H. Naqib*
Department of Physics, University of Rajshahi, Rajshahi 6205, Bangladesh
*Corresponding author, email: salehnaqib@yahoo.com

**Abstract**

In recent times transition metal ternary borides with layered structure have attracted much attention of the materials science community. In this study, we have used density functional theory (DFT) based first-principles investigation of the physical properties of prospective NbAlB compound for the first time. From the analysis of the cohesive energy and enthalpy of formation, it was found that NbAlB is chemically stable. The physical properties of NbAlB have been compared and contrasted with those obtained for MoAlB. Both these MAB phases are elastically anisotropic, mechanically stable, machinable and brittle materials. Structural and elastic features reflect the layered features. The estimated hardness of NbAlB is 19.0 GPa comparable to that of MoAlB (20.8 GPa) suggesting that predicted NbAlB is a hard compound and is suitable for heavy duty industrial applications. NbAlB is more machinable than MoAlB. Electronic band structure calculations reveal conventional metallic behavior with the electronic density of states at the Fermi level arising mainly due to the Nb 4$d$ orbitals in NbAlB. The electronic density of states at the Fermi level is significantly higher in NbAlB in comparison to MoAlB, indicating that NbAlB is expected to exhibit higher level of electrical conductivity. Electronic dispersion is highly anisotropic for both MoAlB and NbAlB with substantially large electronic effective masses in the out-of-plane directions. The bonding features have been elucidated via the analysis of the band structure and charge density distribution. Both the compounds have mixed covalent, ionic and metallic bonding characteristics. The Fermi surfaces of MoAlB and NbAlB consists of electron and hole like sheets. The Debye temperatures of MoAlB and NbAlB are comparable. The estimated melting temperature of NbAlB is somewhat lower than that of MoAlB. NbAlB shows excellent reflection characteristics suitable to be used as an efficient solar reflector. NbAlB is also expected to absorb ultraviolet radiation very effectively. Unlike elastic, bonding, and electronic anisotropy, the optical spectra are fairly isotropic for NbAlB with respect to the polarization of the incident electric field.

## 1. Introduction

Transition-metal borides, carbides, and nitrides have high melting points, high thermal shock resistance, and are among the hardest materials known so far, which are properties required for both heavy duty bulk and coating applications. When combined with a third element, they can retain or form atomically layered structures. For this reason, they are often termed as nanolaminates. One such example is the widely studied $M_{n+1}AX_n$ (MAX) phases (where M is a transition metal, A is an A-group element, and X is C and/or N) of a hexagonal structure with *P63/mmc* symmetry, having a combination of ceramic and metallic attributes [1,2]. These compounds exhibit strongly anisotropic physical properties. The selective removal of the A element, typically Al, upon etching results in the corresponding two-dimensional (2D) material, a so-called MXene phase [3] that has shown an outstanding potential for a wide range of applications, from energy storage [4] to electronics and shielding of electromagnetic radiation [5]. Compounds with similar in-plane chemical order in 3D-layered transition-metal borides have, to this date, not been found. But ternary metallic boride with different crystal symmetry was first synthesized by Ade and Hillebrecht [6] which like the MAX phase compounds also exhibit nanolaminated structure. Due to the similarity in physical properties with the MAX compounds, MAB phase has become an attractive material for investigation. In recent times, Lu et al. [7] carried out the study on structural properties of $AlCr_2B_2$ and $AlFe_2B_2$ compounds which belong to the MAB phase. In particular, some experimental works on the application of the $AlFe_2B_2$ compound as a magnetocaloric material have led to the rise in the interest in these types of structures [8,9]. This compound gained particular attention as a prospective for a rare-earth free magnetocaloric material with a near-room temperature magnetic phase transition at around 304 K. $Mn_2AlB_2$ MAB compound was also experimentally studied to explore its magnetic ground state [10]. The MAB phases have a large palette of structural variations [11,12]. The realized phases are classified as follows: $M_{n+1}AlB_{2n}$ having a single layer of Al with n = 1−3 where M belongs to a transition metal (n = 1 in the space group *Cmmm*: including $Cr_2AlB_2$, $Mn_2AlB_2$, and $Fe_2AlB_2$; n = 2 in the space group *Pmmm*: including $Cr_3AlB_4$; n = 3 in the space group *Cmmm*: $Cr_4AlB_6$) [11,13-15]. Besides, there are MAlB with double layers of Al (MoAlB and WAlB in the space group *Cmcm*) [11,16] and $M_4AlB_4$ with double layers of M atoms ($Cr_4AlB_4$ in the space group *Immm*) [11,17]. It is worth noticing that all structures have orthorhombic symmetry. The chemical diversity of MAB phases has so far been limited to M atoms from group 6 to 8 and A commonly being Al. Effects of alloying has been studied experimentally for $(M_{1-x}Mo_x)AlB$ (space group *Cmcm*) [11,15,16,18] with M = W and Cr and for $Fe_{2-x}Mn_xAlB_2$ [11,19] (space group *Cmmm*). Furthermore, theoretical investigations of the physical properties of $Fe_{2-x}M'_xAlB_2$ (space group *Cmmm*), with M' = Cr, Mn, Ni, Co [11,20] has been carried out.

The crystal structures of MAB compounds depend on the M:B ratio and the number of interleaved planes containing the Al atom. Typically, the MAlB, $M_2AlB_2$, $M_3AlB_4$, $M_4AlB_6$, and $M_4AlB_4$ proto-types are referred to as the 222, 212, 314, 416, and 414 MAB compounds, respectively [11]. Structural units of MAB phases consist of M–B blocks, containing the face-sharing trigonal prisms ($BM_6$), spaced by mono- or bi-layers of Al. At the centre of the $BM_6$ trigonal prisms are the B atoms, which are separated by relatively short distances from each other forming covalently bonded chains. In all cases, $(MB)_{2z}A_x(MB_2)_y$ (with z = 1 - 2; x = 1 - 2; y = 0 - 2) is the generalized chemical formula to describe all these structures [11]. Compared to the large number of MAX phase nanolaminates, the MAB family of compounds is severely limited. Experimental synthesis and theoretical prediction of new MAB compounds with improved physical characteristics are important.

Previously, we have studied the structural, elastic, electronic, optical and bonding properties of experimentally synthesized MoAlB compound via ab-initio technique in details [12]. In this work we have checked the phase and structural stability of the isostructural predicted MAB phase NbAlB and found it to be stable. We have investigated the structural, elastic, electronic, optical, bonding and thermal properties of NbAlB in the ground state using density functional theory based calculations. The physical properties of MoAlB and NbAlB are compared and contrasted. Like MoAlB, the predicted NbAlB possesses a number of attractive physical properties which can be utilized in future engineering and optoelectronic applications.

The rest of this paper has been structured as follows. In section 2 we have given a brief description of the computational methodology. Section 3 contains the computed results and relevant discussions. Finally, in section 4, the important findings are summarized.

## 2. Computational scheme

The computational details for the previously investigated MoAlB can be found elsewhere [12]. The present study for various properties of metallic compound NbAlB is based on *ab-initio* self-consistent field linear combination of atomic orbitals (SCF-LCAO) computer program CASTEP (Cambridge Serial Total Energy Package) with the density functional theory (DFT). The calculations are carried out by computational methods implemented in the CASTEP code [21] which uses the plane-wave pseudopotential based on density functional theory with the generalized gradient approximation (GGA) in the scheme of Perdew-Burke-Ernzerhof (PBE) [22]. In this method the interactions between ion and electron are represented by ultra-soft Vanderbilt-type pseudopotential [23] for Nb, Al and B atoms. The basis sets of the valence electron states are set to be $4d^4\,5s^1$ for Nb, $3s^2\,3p^1$ for Al, and $2s^2\,2p^1$ for B. A ($12\times3\times13$) k-point mesh has been used as the Monkhorst–Pack grid [24] to sample the Brillouin zone (BZ). The plane wave basis set cut-off energy of 600 eV has been used. To get the lowest energy optimized crystal structure of MoAlB, geometry optimization carried out using the Broyden Fletcher Goldfarb Shanno (BFGS) minimization protocol [25]. Geometry optimization is achieved using the convergence thresholds of $2\times10^{-5}$ eV/atom for the total energy, 0.05 eV/Å for the maximum force, 0.1 GPa for the maximum stress and $2\times10^{-3}$ Å for maximum displacement. For charge density mapping, $18\times4\times18$ k-point mesh is used for NbAlB. Electronic band structure is calculated using the optimized crystal structure. For an orthorhombic crystal, there are nine independent elastic constants, i.e. $C_{11}$, $C_{22}$, $C_{33}$, $C_{44}$, $C_{55}$, $C_{66}$, $C_{12}$, $C_{13}$ and $C_{23}$ due to symmetry between the stress and strain tensors. The elastic constants can be determined using the CASTEP code by computing the resulting stress generated due to applying a set of given homogeneous deformation of a finite value. This method has been effectively used to predict the elastic properties of a series of materials including many metallic systems [12,26-28]. The bonding characters in NbAlB and MoAlB are explored with the help of Mulliken population analysis (MPA) [21,29] and Hirshfeld population analysis (HPA) [21,30]. The charge density distributions in various crystallographic planes are also used for the same. The thermophysical parameters are calculated from the elastic constants and moduli. The energy/angular frequency dependent optical parameters are obtained using the matrix elements of photon induced transitions of electrons from the valence band to the conduction band [21]. These matrix elements depend on the electronic band structures calculated for the optimized crystal structures.

## 3. Results and analysis

### 3.1. Structural properties

Crystal structure and symmetry play vital roles in determining many of the properties of a solid, such as elastic constants, electronic band structure, and optical properties. The physical and electronic properties of solids depend entirely upon the arrangement of the atoms that make up the solid and the distances among them. The crystal structures of NbAlB and MoAlB are simple orthorhombic with space group $C_{mcm}$-$D_{2h}^{17}$ (No.63), where all the atoms Nb, Mo, Al, and B occupy the 4c Wyckoff positions. The corresponding point group is *mmm*. The equilibrium lattice parameters of these two MAB compounds are shown in Table 1.

Table 1: The optimized lattice parameters and cell volume (*V*) of NbAlB and MoAlB compounds.

| Phase | $a$ (Å) | $b$ (Å) | $c$ (Å) | $V$ (Å$^3$) | Ref. |
|---|---|---|---|---|---|
| NbAlB | 3.342 | 14.708 | 3.115 | 153.11 | This work |
| MoAlB | 3.215 | 14.049 | 3.106 | 140.28 | [12] |

The lattice parameters and the cell volume of NbAlB are larger than those of MoAlB. This difference is mainly due to the difference in the atomic radii of Nb and Mo. The schematic crystal structure of MoAlB is shown in Fig. 1. The predicted compound NbAlB is completely isostructural to the MoAlB MAB compound.

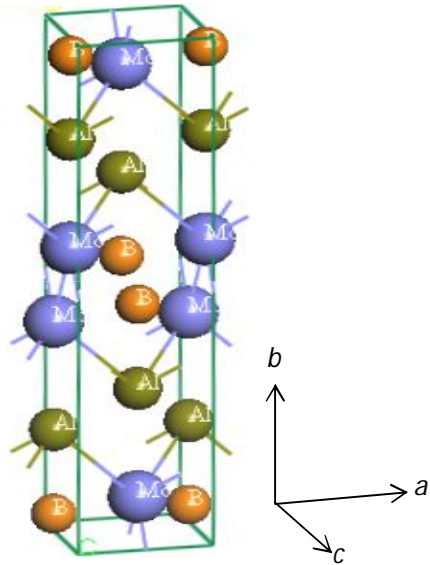

Figure 1: Schematic crystal structure of MoAlB. The crystallographic directions are also shown.

The unit cell of *M*AlB (*M* = Nb, Mo), contains four formula units i.e., it has 12 atoms. Each cell contains four Nb or Mo with four Al atoms and four B atoms.

*3.2. Phase stability*

Before proceeding any further, we studied the phase stability of the predicted compounds NbAlB by calculating the cohesive energy per atom using the following equation:

$$E_{Coh} = E_{NbAlB} - (E_{Nb} + E_{Al} + E_{B}) \quad (1)$$

where, $E_{NbAlB}$ is the total energy per formula unit of the compound and $E_{Nb}$, $E_{Al}$, $E_{B}$ are the total energy of single Nb, Al and B atom, respectively. The calculated value of the cohesive energy per atom is found to be -22.35 eV/atom, for the NbAlB MAB compound. This indicates that the predicted MAB phase should be chemically stable. Again, we examine the stability of NbAlB by calculating the formation of enthalpies (ΔH) from the competing binary phases:

$$\tfrac{1}{2} NbAl_2 + \tfrac{1}{2} NbB_2 \rightarrow NbAlB \quad (2)$$

where, $NbAl_2$ is taken in the P6$_3$/mmc structure and $NbB_2$ is taken in the P6/mmm structure. Using Equation (2), we obtain (ΔH)$_{NbAlB}$ = -172.97 eV/f.u. A negative formation enthalpy of NbAlB means that the predicted compound has phase stability. It should be mentioned that the other MAB phase under consideration, the MoAlB compound, has already been experimentally synthesized and has been proven to be phase stable under ambient conditions.

*3.3. Elastic and mechanical properties*

Elastic properties of solids are directly linked with atomic bonding and cohesive energy. Elastic constants and moduli determine the mechanical behavior of solids completely under applied stress. Various acoustic and thermo-mechanical properties are also related closely to the elastic parameters. The single crystal elastic stiffness coefficients are the proportionality coefficients relating the applied strain to the computed stress. Both stress and strain have three tensile and three shear components, giving six components in total. The linear elastic stiffness $C_{ij}$ form thus a 6×6 symmetric matrix with a maximum of 21 different components, such that $\sigma_i = C_{ij}\varepsilon_j$ for small stresses and strains [31]. Any symmetry present in the crystal structure may make some of these components equal and may let some other components vanish. Mathematically, the second-order independent elastic constants are defined as the second derivatives of the ground state energy with respect to strain components. It can be expressed as:

$$C_{ij} = \frac{1}{V}\left(\frac{\partial^2 E}{\partial e_i \partial e_j}\right) \quad (3)$$

where, *E* and *V* are the energy and volume of the unit cell, respectively.

For an orthorhombic crystal the most common 21 non-zero independent elastic constants reduce to nine components, i.e., $C_{11}$, $C_{22}$, $C_{33}$, $C_{44}$, $C_{55}$, $C_{66}$, $C_{12}$, $C_{13}$ and $C_{23}$ due to symmetry considerations. These nine independent elastic constants determine the elastic behavior of an orthorhombic system completely. These parameters also determine the values of the uniform elastic moduli. The necessary and sufficient criteria for mechanical stability [32,33] of an orthorhombic system are as follows:

$C_{11} > 0; C_{44} > 0; C_{55} > 0; C_{66} > 0;$
$C_{11}C_{22} > C_{12}^2$
$(C_{11}, C_{22}, C_{33} + 2C_{12}C_{13}C_{23}) > (C_{11}C_{23} + C_{22}C_{13} + C_{33}C_{12})$

For estimations of elastic parameters for polycrystalline materials, the Voigt-Reuss-Hill (VRH) approximation [34-36] is widely used. The Voight and Reuss approximations give the upper and lower limits of the polycrystalline elastic moduli, respectively. According to Hill, the actual effective elastic moduli for polycrystals are expressed as the arithmetic mean of the two abovementioned limits – due to Voigt and Reuss:

$B = \frac{1}{2}(B_V + B_R)$ and $G = \frac{1}{2}(G_V + G_R)$.
$$B_V = (1/9)\{(C_{11} + C_{22} + C_{33}) + 2(C_{13} + C_{12} + C_{23})\}$$
$$G_V = (1/15)\{C_{11} + C_{12} + C_{33} + 3(C_{44} + C_{55} + C_{66}) - 4(C_{12} + C_{13} + C_{23})\}$$
$$B_R = 1/\{(S_{11} + S_{22} + C_{33}) + 2(C_{13} + C_{12} + C_{23})\}$$
$$G_R = (1/15)\{S_{11} + S_{12} + S_{33} + 3(S_{44} + S_{55} + S_{66}) - 4(S_{12} + S_{13} + S_{23})\}$$

Further, the calculated bulk moduli $B$ and shear moduli $G$ allow us to obtain the Young's moduli $Y$ and the Poisson's ratio $\nu$ as:

$$Y = \frac{9BG}{3B + G}$$
$$\nu = \frac{3B - 2G}{2(3B + G)}$$

The computed single crystal elastic constants are given in Table 2 below.

Table 2: Single crystal elastic constants (all in GPa) of predicted NbAlB and already synthesized MoAlB MAB compounds.

| Phase | $C_{11}$ | $C_{22}$ | $C_{33}$ | $C_{44}$ | $C_{55}$ | $C_{66}$ | $C_{12}$ | $C_{13}$ | $C_{23}$ | Ref. |
|---|---|---|---|---|---|---|---|---|---|---|
| NbAlB | 269.4 | 265.7 | 352.2 | 155.5 | 168.7 | 178.9 | 128.9 | 143.4 | 106.0 | This work |
| MoAlB | 349.1 | 320.2 | 399.6 | 190.3 | 160.0 | 169.0 | 141.8 | 146.1 | 118.2 | [12] |

Among the nine independent elastic constants, $C_{11}$, $C_{22}$, and $C_{33}$ measure the resistance to linear strain along the crystallographic a-, b-, and c-axes, respectively. For both the compounds, $C_{33} > (C_{11}, C_{22})$ indicating that the atomic bonding strengths along the c-direction is strongest in these MAB compounds. The off-diagonal, shear components of the elastic constants are $C_{12}$, $C_{13}$, and $C_{23}$. The elastic tensors $C_{12}$ and $C_{13}$ combine a functional stress component in the crystallographic a-direction with a uniaxial strain along the crystallographic b- and c-axes, respectively. $C_{44}$, on the other hand indicates material's ability of resisting the shear deformation in (100) plane and, $C_{55}$ and $C_{66}$ reflect the resistance to shear in the <011> and <110> directions, respectively. It is found that most of the elastic constants including $C_{44}$ are higher for MoAlB compared to those for NbAlB. Hardness of a solid depends on the value of $C_{44}$, therefore, it is predicted that MoAlB should be harder than NbAlB.

The polycrystalline elastic moduli and some other useful elastic indicators are shown in Table 3.

Table 3: Polycrystalline bulk modulus $B_v$, $B_R$, $B$ (in GPa), shear moduli $G_V$, $G_R$, $G$ (in GPa), Young modulus $Y$ (in GPa), Pugh's ratio $B/G$ and Poisson's ratio $v$, Cauchy pressure $C_p$ (in GPa), tetragonal shear modulus, $C_t$ (in GPa) for the NbAlB and MoAlB MAB phases in the ground state.

| Phase | $B_V$ | $B_R$ | $B$ | $G_V$ | $G_R$ | $G$ | $Y$ | $B/G$ | $v$ | $C_p$ | $C_t$ | Ref. |
|---|---|---|---|---|---|---|---|---|---|---|---|---|
| NbAlB | 182.7 | 179.9 | 181.3 | 134.6 | 116.6 | 125.6 | 306.1 | 1.44 | 0.22 | -49.5 | 70.25 | This work |
| MoAlB | 209.0 | 207.5 | 208.0 | 148.2 | 139.1 | 144.6 | 351.9 | 1.44 | 0.22 | -72.1 | 103.65 | [12] |

It is noticed from Table 3 that the difference between $B_V$ and $B_R$ as well as $G_V$ and $G_R$ are quite small. According to Hill, the difference between these limiting values should be proportional to the degree of elastic anisotropy of the solid [36]. Hence, the ternary borides under study should exhibit small anisotropy in elastic behavior. Bulk modulus reflects the ability of solid to resist compression due to uniform hydrostatic pressure. The shear modulus, on the other hand, reflects the ability of the solid to resist shape changing external stress. For both the MAB phases $B > G$, implying that the mechanical failure of NbAlB and MoAlB should be controlled by the shearing stress rather than the volume stress. The Young modulus quantifies the resistance of the solid to length changing tensile stress. This parameter determines the stiffness of a solid. From the values of $Y$ of the two compounds, it may be inferred that MoAlB is significantly stiffer than NbAlB. A solid is either brittle or ductile for most practical situations. The bulk modulus to shear modulus ratio, $B/G$, known as the Pugh's ratio [37] is frequently used as a measure of brittle or ductile behavior of solids. The larger the $B/G$ value, the higher the ductility of solids. The critical value of $B/G$ is 1.75. Specifically, if $B/G > 1.75$, the material should be characterized as ductile, otherwise the material should be brittle. From Table 3, it was noted that both MoAlB and NbAlB should behave in a brittle manner. The parameter which can predict the stability of crystal systems against shear, is the Poisson's ratio. The relatively small Poisson's ratio is related to the systems stability against shear. Therefore, it is expected that NbAlB and MoAlB should be stable against shear due to their low Poisson's ratio $v$. The Poisson's ratio also predicts the character of inter-atomic forces and brittleness/ductility in solids [38]. The critical value of $v$ is 0.26 [38,39]. To be specific, the Poisson's ratio is greater than 0.26 for ductile solids and it is less than 0.26 for brittle materials. From table 3 we observe that the Poisson's ratios of NbAlB and MoAlB are less than 0.26. So, we suggest that NbAlB and MoAlB are brittle compounds. There is another elastic parameter which can be used to judge brittleness/ductility; the Cauchy pressure [$C_p = (C_{23} - C_{44})$] [40]. A negative value of the Cauchy pressure implies brittleness and presence of significant angular bonding. Both the MAB compounds have significantly negative Cauchy pressure reaffirming their brittle character. High and negative values of $C_p$ further suggest that these compounds should have significant (angular) covalent bonding between the atomic species. The tetragonal shear modulus, $C_t$, is calculated from, $C_t = (C_{11} - C_{12})/2$ [41]. This parameter is closely linked with the transverse acoustic wave velocity in a crystal and its dynamic stability [41]. Large positive values of $C_t$ are indicative of dynamic stability of NbAlB and MoAlB.

There are several other important mechanical parameters of solids in view of their engineering applications e.g., the Vickers hardness ($H_V$), machinability index ($\mu_M$), and the fracture toughness ($F_T$). We have calculated all these parameters using the formulae given below [42-45]:

$$H_V = 2\left[\left(\frac{G}{B}\right)^2 G\right]^{0.585} - 3 \qquad (4)$$

$$\mu_M = B/C_{44} \qquad (5)$$

$$F_T = V_o^{(1/6)} G (B/G)^{1/2} \qquad (6)$$

In Eqn. 6, $V_0$ is the volume per atom of the compound. The computed values of $H_V$, machinability index $\mu_M$, and the fracture toughness $F_T$ of the MAB compounds under study are given in Table 4.

Table 4: The Vickers hardness, machinability index, and fracture toughness of NbAlB and MoAlB compounds.

| Phase | $H_V$ (GPa) | $\mu_M$ | $F_T$ (MPa.m$^{1/2}$) | Ref. |
|---|---|---|---|---|
| NbAlB | 19.01 | 1.167 | 2.613 | This work |
| MoAlB | 20.81 | 1.095 | 2.303 | [12] |

The computed values of Vickers hardness shows that both the compounds are fairly hard; harder than most of the MAX phase compounds including the recently discovered ternary boride phases [46,47]. Hardness of MoAlB is almost 10% higher than that of NbAlB, consistent with the values of other elastic parameters. The machinability indices are moderate. The machinability index is a measure of the ease with which a solid can be cut into different shapes. It also gives a measure of dry lubricity. Materials with high machinability index exhibit high level of dry lubricity. The fracture toughness quantifies the ability of a solid to withstand high level of mechanical stress. The fracture toughness values of NbAlB and MoAlB compounds are quite high; higher than many of the MAX phase borides. High hardness and high fracture toughness of these two compounds make them suitable for heavy duty engineering applications.

## 3.4. Elastic anisotropy

Most of the crystalline solids are elastically anisotropic. This leads to mechanical anisotropy which is important from the application's point of view. The elastic anisotropy has a significant effect on large number of physical processes, including the plastic deformation in solids, the development of cracks and its propagation, micro-scale cracking in ceramics, unconventional phonon modes, dislocation dynamics, internal abrasion, etc [48]. Various formalisms are used to examine the anisotropic properties of crystals. In this section, we have calculated the shear anisotropy factors $A_1$, $A_2$ and $A_3$, percentage anisotropy factors in bulk and shear moduli, $A_B$ and $A_G$, respectively, and the universal anisotropy index $A^U$ for the NbAlB and MoAlB compounds. The formulae used to calculate these anisotropy indices are given below [48,49]:

$$A_1 = \frac{\frac{1}{6}(C_{11} + C_{12} + 2C_{33} - 4C_{13})}{C_{44}}$$

$$A_2 = \frac{2C_{44}}{C_{11} - C_{12}}$$

$$A_3 = A_1 \cdot A_2 = \frac{\frac{1}{3}(C_{11} + C_{12} + 2C_{33} - 4C_{13})}{C_{11} - C_{12}}$$

$$A_B = \frac{B_V - B_R}{B_V + B_R} \times 100\%$$

$$A_G = \frac{G_V - G_R}{G_V + G_R} \times 100\%$$

$$A^U = 5\frac{G_V}{G_R} + \frac{B_V}{B_R} - 6 \geq 0$$

The calculated anisotropy indices are summarized in Table 5.

Table 5: The shear anisotropy factors $A_1$, $A_2$, $A_3$, percentage anisotropy factors $A_B$ and $A_G$, and the universal anisotropy index $A^U$ of NbAlB and MoAlB MAB compounds.

| Phase | $A_1$ | $A_2$ | $A_3$ | $A_B$ (%) | $A_G$ (%) | $A^U$ | Ref. |
|---|---|---|---|---|---|---|---|
| NbAlB | 1.86 | 1.66 | 2.58 | 0.77 | 7.19 | 0.793 | This work |
| MoAlB | 1.67 | 1.33 | 1.75 | 0.48 | 3.14 | 0.333 | [12] |

The shear anisotropy factors $A_1$, $A_2$ and $A_3$ indicate the level of elastic anisotropy for the {100} shear planes between the ⟨011⟩ and ⟨010⟩ directions, for the {010} shear plane between ⟨101⟩ and ⟨001⟩ directions, and for the {001} shear planes between ⟨110⟩ and ⟨010⟩ directions, respectively. For a crystal which is isotropic with respect to the different shear planes, $A_1 = A_2 = A_3 = 1$. Departure from unity quantifies the anisotropy in shear. It is observed that both the MAB phases are anisotropic for shape deforming shearing stress. The percentage anisotropy in the bulk modulus is very low for NbAlB and MoAlB. The percentage anisotropy in the modulus of rigidity is moderate. The universal anisotropy index, applicable to crystals with any symmetry [50], also suggests that NbAlB and MoAlB are moderately anisotropic. Overall, the anisotropy level is greater in the predicted NbAlB compound in comparison to MoAlB. This is probably due to relatively weaker chemical bondings along the *c*-direction in NbAlB.

*3.5. Electronic band structure and energy density of states*
The calculation of the electronic band structure helps one to understand the shape of the Fermi surface. The charge transport and optical properties of material can be understood from the character of dominant bands near the Fermi level. Fig. 2 presents the energy band structure for the optimized crystal structure of NbAlB along the high symmetry directions (*Γ - Z - T - Y - S - X - U - R*) of the first BZ in the energy range from −12 to +5 eV. The Fermi level is chosen to be the zero of the energy scale. For NbAlB, the valence and conduction bands are seen to overlap, thus indicating metallic behavior. The valence band and conduction band profiles of NbAlB are fairly similar to those of the previously studied MoAlB compound [12].

The total number of electronic bands for NbAlB is 95. The bands which cross the Fermi level are shown (colored) in Fig. 2 including their band numbers. Most of the bands crossing the Fermi level show hole-like character (Fig. 2). These bands also show varying degree of dispersion. For example, band 36 is fairly dispersive; on the other hand, band 38 is less dispersive. This implies that the charge carriers in different electronic orbitals have different effective masses and mobilities. Overall, the degree of dispersion of the bands running along the crystallographic *c*-axis is very low (bands 38 and 40). These nearly nondispersive bands will make the effective masses of very high for the charge carriers moving in the *c*-direction and will make the electronic transport properties highly anisotropic.

Fig. 3 shows the electronic energy density of states of NbAlB. The electronic energy density of states of MoAlB was analyzed in our previous work [12]. The electronic energy density of states (DOS) of a system gives the number of electronic states per unit interval of energy at each energy level that is available to be occupied. The DOS governs many electronic, optical, and bonding properties and consequently plays an important role in solid state physics. A high DOS at a specific energy level means that there are many states available for occupation. A DOS of zero value means that no states can be occupied at that particular energy level. The total and partial densities of states at the Fermi level, TDOS and PDOS, respectively, of NbAlB are shown in Fig. 3. To understand the contribution of each atomic orbital to the total density of states, we have calculated the PDOS of Nb, Al, B in NbAlB. The non zero values of TDOS at the Fermi level is the evidence that like MoAlB [12], NbAlB should also exhibit metallic behavior. At the Fermi energy ($E_F$), the value of TDOS for NbAlB is 5.45 states/eV-unit cell. The corresponding value for the MoAlB compound is 2.90 states/eV-unit cell. Therefore, the TDOS of the predicted MAB compound NbAlB is significantly larger than that of MoAlB. We, therefore, predict that the electrical conductivity of NbAlB should be much higher than that of MoAlB.

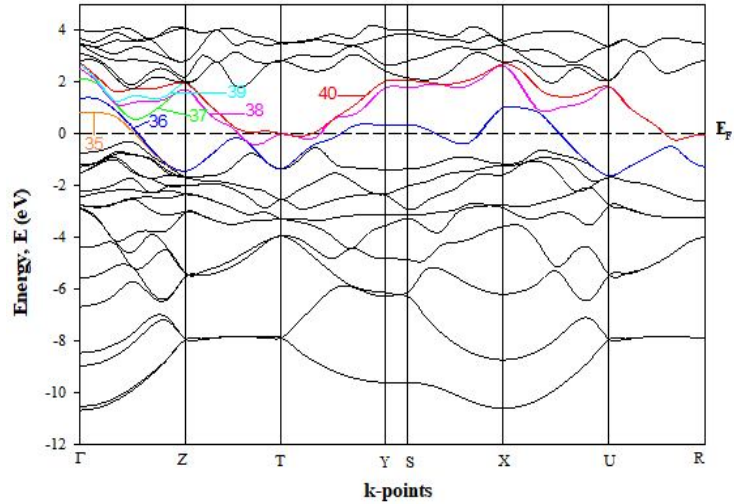

Figure 2: Electronic band structure of NbAlB. The horizontal dashed line marks the Fermi energy.

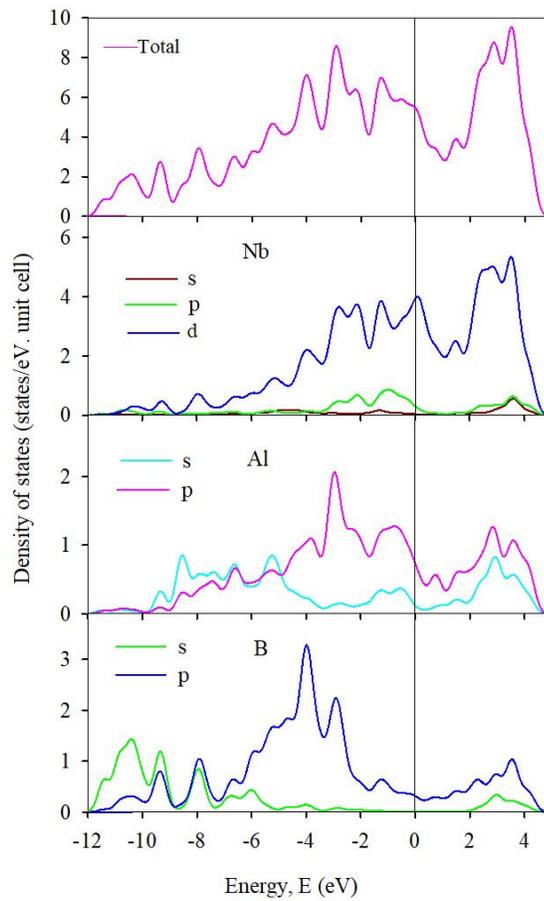

Figure 3: The TDOS and PDOS of NbAlB. The vertical dashed line at zero energy marks the Fermi level.

The TDOS and PDOS of NbAlB show many interesting features. The TDOS at the Fermi level arises mainly from the contribution of the Nb 4*d* electronic states. There is some contribution from the Al 3*p* and B 2*p* orbitals as well. There is significant hybridization among the Nb 4*d*, Al 3*p*, and B 2*p* electronic states close to the Fermi energy. Such strong hybridization leads to the formation of covalent bondings between the orbitals involved [51,52]. The conduction band of NbAlB is dominated by the Nb 4*d* electronic states. The lower energy part of the valence band, i.e. between -12 eV to –9.33 eV, the B 2*s* states contribute much to the TDOS. In the energy range from -7.5 eV to -3.39 eV Nb 4*d*, Al 3*p* as well as B 2*p* contribute significantly. Finally, the valence band from -3.39 to Fermi level is dominated mainly by Nb 4*d* states with a small contribution of Al 3*p* and B 2*p* orbitals. For the existing MAB compound MoAlB, there is a pseudogap near the Fermi level [12] which suppresses the TDOS and at the same time it serves as the borderline between the bonding and anti-bonding electronic states and is cooperative for stabilizing the structure of MoAlB.

*3.6. Charge density distribution*

Since all the mechanical and elastic properties of solids are related to the nature and strength of the atomic bonding, it is important to explore it in greater detail. Investigation of the charge distribution in various crystal planes gives useful insights regarding the chemical bonding in a solid. The electron charge density map reveals the electron densities associated with the chemical bonds. It includes areas of positive as well as negative charge densities and indicates both accumulation and depletion of electronic charges. From the charge density map, the covalent bonds can be identified via the accumulation of charges between two atoms. The existence of ionic bonds can be predicted from a negative and positive charge balance at the atom positions. The valence electronic charge density map (in the units of e/Å$^3$) for NbAlB is shown in Fig. 4 in the (110) and (011) crystallographic planes. The scale on right hand side of charge density maps shows the intensity of total electron density. The blue color shows high charge (electron) density and red color shows low charge (electron) density.

From Fig. 4, it is observed that Nb contains the highest charge density and Al contains the lowest charge density in the (011) plane. The electronic charge density distributions are different for the (110) and (011) planes. Therefore, the charge density distributions in NbAlB are anisotropic. Moreover, the unequal charge distributions surrounding the atoms are not circular. There are direction dependences. This implies that there are both ionic and covalent contributions to the atomic bondings. Uniform charge background also indicates that there are some metallic bondings in NbAlB. The overall charge distribution in NbAlB is qualitatively similar to that of MoAlB [12] where the Mo atom has the highest charge density.

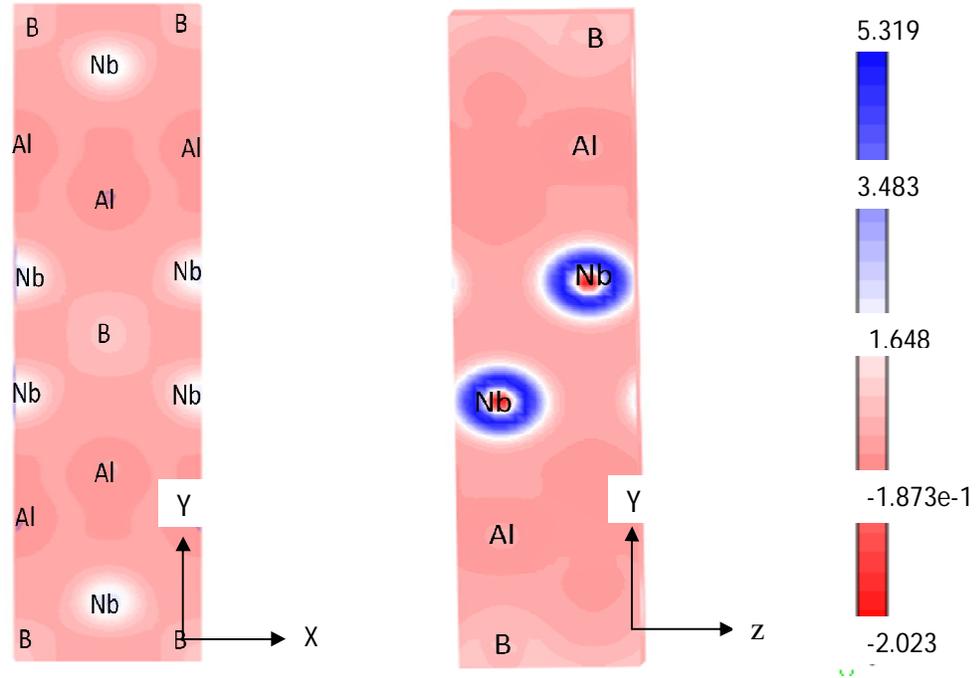

Figure 4: The electronic charge density map for NbAlB in the (110) and (011) plane.

*3.7. Mulliken atomic and bond overlap population*
The Mulliken bond populations are calculated to understand the bonding behavior in greater detail. The results obtained from the calculations for NbAlB are given in Tables 6 and 7. The earlier results [12] obtained for the MoAlB MAB compound is also displayed in Tables 6 and 7 for comparison. The Mulliken bond populations measure the degree of overlap of the electron clouds of the two bonding atoms, and its highest and lowest values signify the strength of covalency and ionicity, respectively. The total overlap population for any pair of atoms in a molecule is in general made up of positive and negative contributions. If the total bond overlap population is positive then they are bonded; if negative, then they are anti-bonded [29].

For the predicted MAB phase NbAlB, Nb and Al contain positive Mulliken charge and B contains negative Mulliken charge so electrons are transferred from the Nb and Al atoms to the B atom, whose values are +0.37e and +0.19e respectively. This suggests that an ionic contribution to the bonding is present in this compound. Thus, we can say Nb-B and Al-B bonds have partially ionic nature in NbAlB. From the values of effective valence charge (EVC), which is defined as the difference between the formal ionic charge and the Mulliken charge on the cation species, the degree of covalency and/or ionicity may be inferred [53]. It determines the strength of a bond either as covalent or ionic. When the effective valence charge has exactly zero value, an ideal ionic bond occurs. An atom associated with a non-zero effective valence charge indicates covalent bonding and by the degree of deviation from zero, the level of covalency can be estimated. Table 6 lists the calculated effective valence charges that indicate the existence of covalency in chemical bonding inside NbAlB and MoAlB. The order of covalency level of the chemical bonds for NbAlB can be expressed as:

B-B > Nb-Al > Al-Al > Al-B > Nb-B. So we can say that, Nb-Nb bond is more covalent than any other bonds in NbAlB. Table 6 also reveals that the order of covalency level found in chemical bonds in MoAlB can be expressed as [12]: B-B > Al-Al > Al-Mo > B-Mo > Al-B. The reverse of these orders should indicate the order of ionicity level of the chemical bonds in NbAlB and MoAlB.

Table 6: Mulliken atomic charge populations of NbAlB and MoAlB MAB compounds.

| Phase | Atoms | s | p | d | Total charge | Mulliken Charge | EVC |
|---|---|---|---|---|---|---|---|
| NbAlB | B | 0.96 | 2.60 | 0.00 | 3.56 | -0.56 | -- |
| | Al | 0.89 | 1.93 | 0.00 | 2.81 | 0.19 | 2.81 |
| | Nb | 2.08 | 6.41 | 4.14 | 12.63 | 0.37 | 4.63 |
| MoAlB | B | 0.95 | 2.53 | 0.00 | 3.49 | -0.49 | -- |
| | Al | 0.85 | 1.87 | 0.00 | 2.72 | 0.28 | 2.72 |
| | Mo | 2.11 | 6.47 | 5.21 | 13.79 | 0.21 | 5.79 |

Table 7: Results of Mulliken bond population analysis of NbAlB and MoAlB compounds.

| Phase | Bond | Number of bonds | Bond length (Å) | Bond overlap population |
|---|---|---|---|---|
| NbAlB | B-B | 2 | 1.770 | 1.521 |
| | B-Al | 4 | 2.405 | 0.132 |
| | B-Nb | 4 | 2.368 | 0.564 |
| | Al-Al | 2 | 2.609 | 1.014 |
| | Al-Nb | 4 | 2.730 | 0.596 |
| MoAlB | B–B | 2 | 1.809 | 1.384 |
| | B–Al | 4 | 2.320 | 0.153 |
| | B–Mo | 4 | 2.369 | 0.665 |
| | Al–Al | 2 | 2.667 | 0.951 |
| | Al–Mo | 4 | 2.711 | 0.737 |

Table 7 reveals that for both the MAB phases the B-B bond length is the smallest and has the largest bond overlap population. This implies that the overall mechanical strength of the synthesized MAB compound MoAlB is due to the the B-B atomic bondings. The same is true for the predicted NbAlB MAB compound.

*3.8. Thermo-mechanical properties*

Debye temperature, ($\Theta_D$) is one of the simplest and most useful parameters to understand thermodynamic properties of solids. For example, specific heat, stability of lattice, melting point, phonon thermal conductivity etc. depend strongly on the Debye temperature. Physically, Debye temperature signifies the temperature at which all the modes of lattice vibrations are activated. At low temperatures, below the Debye temperature, the quantum mechanical nature of lattice vibrational spectra is manifested.

The Debye temperature can be calculated using the mean sound velocity as follows [54]:

$$\Theta_D = \frac{h}{k_B}\left[\frac{3n}{4\pi V_0}\right]^{1/3} v_a \qquad (7)$$

where, $h$ is Planck's constant, $k_B$ is the Boltzmann's constant, $V_0$ is the volume of the optimized unit cell and $n$ is the number of atoms in the unit cell. The average sound velocity $v_a$ is given by [54]:

$$v_a = \left[\frac{1}{3}\left(\frac{2}{v_t^3} + \frac{1}{v_l^3}\right)\right]^{-1/3}$$

where, $v_l$ and $v_t$ are the longitudinal and transverse sound velocities, respectively, obtained using the shear modulus $G$, the bulk modulus $B$ and density $\rho$ of the solid [54].

$$v_l = \left(\frac{3B+4G}{3\rho}\right)^{1/2}$$

and

$$v_t = \left(\frac{G}{\rho}\right)^{\frac{1}{2}}$$

The computed values of the Debye temperatures and sound velocities of NbAlB and MoAlB are displayed in Table 8.

Table 8: Calculated crystal density ($\rho$ in g/cm³), longitudinal, transverse and average sound velocities ($v_l$, $v_t$ and $v_a$ in km/s) and Debye temperature ($\Theta_D$ in K) of NbAlB and MoAlB.

| Compounds | $\rho$ | $v_l$ | $v_t$ | $v_a$ | $\Theta_D$ | Ref. |
|---|---|---|---|---|---|---|
| NbAlB | 5.67 | 7.84 | 4.71 | 5.21 | 664 | This work |
| MoAlB | 6.33 | 7.95 | 4.77 | 5.28 | 693 | [12] |

The acoustic properties and the Debye temperatures of NbAlB and MoAlB are very close together; implying that thermal properties of these two MAB compounds should be quite similar.

We have also estimated the melting temperature, $T_m$, of NbAlB and MoAlB compounds following the relation put forward by Fine et al. [55]:

$$T_m = 354 + 1.5(2C_{11} + C_{33}) \tag{8}$$

Melting temperature is an important thermophysical parameter since it determines the limit for high temperature application of a solid. The calculated values of the melting temperatures of NbAlB and MoAlB compounds are 1690.5 K and 2000.7 K, respectively. Higher melting point of MoAlB suggests that the overall bonding strength among the atoms is stronger in this compound compared to the predicted NbAlB MAB phase.

The lattice anharmonic effect of a compound is determined by the Grüneisen parameter. The Grüneisen parameter, $\gamma$, is an important quantity in the thermodynamics and lattice dynamics because it is related with bulk modulus, heat capacity, thermal expansion coefficient and volume of the solid. A high value of the Grüneisen parameter implies high level of anharmonicity. The Grüneisen parameter can be evaluated from the Poisson's ratio of a solid as follows [56]:

$$\gamma = \frac{3[1+s]}{2[2-3s]} \tag{9}$$

A high value of the Grüneisen parameter results in high crystal compressibility and high coefficient of thermal expansion. The calculated values of the Grüneisen parameter of NbAlB and MoAlB compounds are identical, 1.365, since the Poisson's ratio of these two MAB phases are equal. The Grüneisen parameter of many of the MAX phase compounds have similar moderate values [57,58] indicating moderate levels of lattice anharmonicity and low thermal conductivity at elevated temperatures.

*3.9. Optical properties*

Thorough understanding of the optical properties of materials is essential for the advancement of optical technology and their applications. All the important photon energy dependent optical parameters are controlled by the underlying electronic band structure of the system of interest. In this section we have presented the results of optical properties calculations in details. From the computed complex dielectric constants of NbAlB, all the other optical parameters including the complex refractive index, absorption coefficient, reflectivity, optical conductivity, and loss function were obtained. The details regarding the theoretical formalisms can be found elsewhere [21,59]. We have used 0.5 eV Gaussian smearing in all the calculations. An empirical Drude term with a plasma frequency of 10 eV and damping of 0.05 eV was used for our calculations. To explore the optical anisotropy, electromagnetic waves of different energies were considered with [100] and [001] polarization directions of the electric field vectors. The optical spectra are shown in Figs. 5 below in the energy range from 0 to 25 eV. To the best of our knowledge, comprehensive theoretical study of optical parameters of MoAlB does not exist in the literature. Therefore, no comparison of the optical properties of NbAlB and MoAlB can be made.

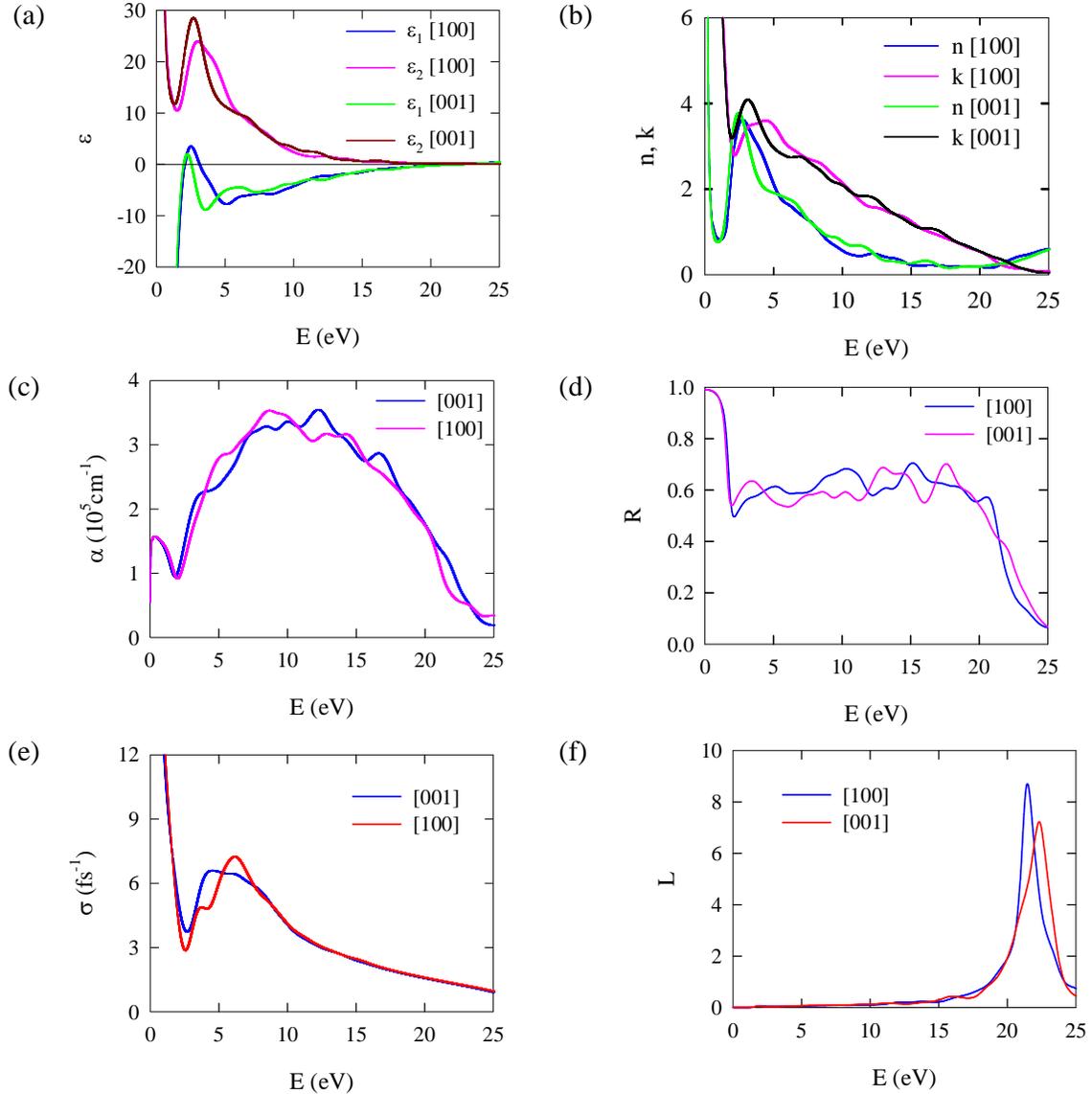

Figure 5: The photon energy dependent (a) real and imaginary parts of the dielectric function, (b) real and imaginary parts of refractive index, (c) absorption coefficient, (d) reflectivity, (e) optical conductivity, and (f) loss function of NbAlB for two different directions of electric field vectors.

The real [$\varepsilon_1(\omega)$] and imaginary parts [$\varepsilon_2(\omega)$] of the dielectric function of NbAlB for the [100] and [001] polarization directions are shown in Fig. 5a. The real part of the dielectric function is related to electrical polarization and anomalous dispersion, while the imaginary part is associated with dissipation of energy of the electromagnetic wave within the medium. The dielectric function of NbAlB exhibits metallic characteristics in the energy ranges for which $\varepsilon_1(\omega) < 0$. The magnitudes of both $\varepsilon_1(\omega)$ and $\varepsilon_2(\omega)$ decrease monotonically in the high energy region. The imaginary part approaches zero around 20 eV. The peaks in the $\varepsilon_2(\omega)$ result from the combined effects of matrix elements of photon induced electronic transitions between electronic states and their respective energy density of states. The highest peaks in both $\varepsilon_1(\omega)$ and $\varepsilon_2(\omega)$ appear in the energy range 2.5 eV – 3 eV in the visible range. There is moderate optical anisotropy in the dielectric function with respect to electric field polarization.

The refractive index, *n* (real part) and extinction coefficients, *k* (imaginary part) of NbAlB for the [100] and [001] polarization directions are displayed in Fig. 5b. The real part of refractive index represents phase velocity of the electromagnetic wave at different energies while the imaginary part, *k*, indicates the attenuation of electromagnetic wave when it passes through a media. The extinction coefficient is peaked at 3.86 eV in the near ultraviolet (UV) region. The low energy value of *n* is quite high. High refractive index solids are suitable for wave guide applications. The optical anisotropy in refractive index is quite low.

The absorption coefficient is related to the optimum solar energy conversion efficiency of solar cell and it determines how far light of a specific energy (wavelength/frequency) can penetrate into the material before being absorbed. The absorption spectrum of NbAlB is shown in Fig. 5c. Finite absorption coefficient at low energy reaffirms the metallic character of NbAlB. The absorption coefficient is quite high in the UV region and peaks at ~10 eV. Absorption coefficient falls sharply at around 20 eV, close to the plasma edge. The optical anisotropy in absorption coefficient is moderate.

The reflectivity is defined as the ratio of the energy of a wave reflected from a surface to the energy possessed by the wave striking the surface. The reflectivity spectra as a function of photon energy of NbAlB phase is shown in Fig. 5d. The reflectivity in the infrared region is very high. For the reflectivity spectrum encompassing the visible to ultraviolet radiation (2 eV – 20 eV) is high and non-selective. This is an attractive optical feature of NbAlB and this predicted MAB phase has high potential to be used as a very efficient reflector of solar radiation. Like the absorption coefficient, the reflectivity of NbAlB decreases sharply at energies above 20 eV. The optical anisotropy is low as far as the reflectivity of NbAlB is concerned.

Since NbAlB has no band gap as evident from the electronic band structure, the photoconductivity starts with zero photon energy as shown in Fig. 5e. The peaks in the optical conductivity show electric field polarization direction dependence. Non-zero photoconductivity at low photon energy once again suggests metallic electronic band structure of NbAlB.

The loss function, $L(\omega)$, for the NbAlB phase under study is displayed in Fig. 5.f. This particular parameter describes the energy loss of a fast electron via exciting collective charge oscillation modes while traversing a material [21,59]. Its peak energy is defined as the bulk plasma energy and the corresponding frequency as the plasma frequency, $\omega_P$. This particular energy/frequency is at the onset of $\varepsilon_2 < 1$ and $\varepsilon_1 = 0$. For NbAlB the highest peaks of the loss function are located at 21.50 eV and 22.53 eV for the [100] and [001] electric field polarization directions, respectively. There is optical anisotropy in the loss peak spectra.

When the incident photon frequency exceeds $\omega_P$, the material becomes transparent to the incident light.

## 4. Conclusions

In this study we have predicted a new member of the technologically prominent MAB phase compound NbAlB, using DFT based computations. The compound is found to be chemically stable. It is also found to be elastically stable. Calculations of elastic constants and moduli show mechanical anisotropy. The physical properties of the predicted NbAlB have been compared to those of already synthesized MAB compound MoAlB [12]. The elastic properties, machinability index, hardness, and Debye temperature of NbAlB are quite close in values of MoAlB compound. The level of elastic/mechanical anisotropy is higher in NbAlB. Both these MAB compounds are hard and brittle in nature. The chemical bonding of NbAlB is dominated by ionic and covalent contributions. Like in the synthesized MoAlB compound, the electronic band structure of NbAlB reveals clear metallic character. We expect higher electrical conductivity of NbAlB due to its higher electronic energy density of states at the Fermi level. The acoustic velocities in NbAlB and MoAlB are comparable. The estimated melting temperature of MoAlB, on the other hand, is significantly higher than that of NbAlB. The optical properties of NbAlB are investigated in detail. All the optical parameters show metallic character. There is optical anisotropy. The compound is found to be a very good absorber of UV radiation. The reflectivity spectrum is nonselective for a wide energy range. NbAlB shows excellent reflection features suitable for reducing solar heating. High value of the real part of the refractive index in the infrared and visible range suggests that the predicted MAB compound NbAlB can be used in the optoelectronic device sector.

**Acknowledgements**

S. H. N. acknowledges the research grant (1151/5/52/RU/Science-07/19-20) from the Faculty of Science, University of Rajshahi, Bangladesh, which partly supported this work.

**Data availability**

The data sets generated and/or analyzed in this study are available from the corresponding author on reasonable request.

**Declaration of interest**

The authors declare that they have no known competing financial interests or personal relationships that could have appeared to influence the work reported in this paper.

**CRediT author statement**

**Mst. Bina Aktar:** Methodology, Formal analysis, Writing-Original draft. **F. Parvin:** Supervision, Writing-Reviewing and Editing. **A. K. M. Azharul Islam:** Supervision, Writing- Reviewing and Editing. **S. H. Naqib:** Conceptualization, Formal analysis, Supervision, Writing- Reviewing and Editing.